\documentstyle[prl,epsf,aps,floats,amssymb]{revtex}  % use this for PRL mode
\begin{document}
\def\etal{{\sl et al.}}                 %et al. - no proceeding comma
\lefthyphenmin=2
\righthyphenmin=3
%------------------------------------------------------------------------
\title{ Ratios of Multijet Cross Sections in
$p\bar{p}$ Collisions at $\surd{s} = 1.8$ TeV}
% LIST_OF_AUTHORS.TEX                 8/10/00            
%
\author{                                                                      
%% names begin here                                                           
B.~Abbott,$^{50}$                                                             
M.~Abolins,$^{47}$                                                            
V.~Abramov,$^{23}$                                                            
B.S.~Acharya,$^{15}$                                                          
D.L.~Adams,$^{57}$                                                            
M.~Adams,$^{34}$                                                              
G.A.~Alves,$^{2}$                                                             
N.~Amos,$^{46}$                                                               
E.W.~Anderson,$^{39}$                                                         
M.M.~Baarmand,$^{52}$
% ...
\\                                                                            
\vskip 0.30cm                                                                 
\centerline{(D\O\ Collaboration)}                                             
\vskip 0.30cm                                                                 
}                                                                             
\address{                                                                     
\centerline{$^{1}$Universidad de Buenos Aires, Buenos Aires, Argentina}       
\centerline{$^{2}$LAFEX, Centro Brasileiro de Pesquisas F{\'\i}sicas,         
                  Rio de Janeiro, Brazil}                                     
\centerline{$^{3}$Universidade do Estado do Rio de Janeiro,                   
                  Rio de Janeiro, Brazil}                                     
\centerline{$^{4}$Institute of High Energy Physics, Beijing,                  
                  People's Republic of China}                                 
\centerline{$^{5}$Universidad de los Andes, Bogot\'{a}, Colombia}             
\centerline{$^{6}$Charles University, Prague, Czech Republic}                 
\centerline{$^{7}$Institute of Physics, Academy of Sciences, Prague,          
                  Czech Republic}                                             
\centerline{$^{8}$Universidad San Francisco de Quito, Quito, Ecuador}         
\centerline{$^{9}$Institut des Sciences Nucl\'eaires, IN2P3-CNRS,             
                  Universite de Grenoble 1, Grenoble, France}                 
\centerline{$^{10}$CPPM, IN2P3-CNRS, Universit\'e de la M\'editerran\'ee,     
                  Marseille, France}                                          
\centerline{$^{11}$LPNHE, Universit\'es Paris VI and VII, IN2P3-CNRS,         
                  Paris, France}                                              
\centerline{$^{12}$DAPNIA/Service de Physique des Particules, CEA, Saclay,    
                  France}                                                     
\centerline{$^{13}$Panjab University, Chandigarh, India}                      
\centerline{$^{14}$Delhi University, Delhi, India}                            
\centerline{$^{15}$Tata Institute of Fundamental Research, Mumbai, India}     
\centerline{$^{16}$Seoul National University, Seoul, Korea}                   
\centerline{$^{17}$CINVESTAV, Mexico City, Mexico}                            
\centerline{$^{18}$FOM-Institute NIKHEF and University of                     
                  Amsterdam/NIKHEF, Amsterdam, The Netherlands}               
\centerline{$^{19}$University of Nijmegen/NIKHEF, Nijmegen, The               
                  Netherlands}                                                
\centerline{$^{20}$Institute of Nuclear Physics, Krak\'ow, Poland}            
\centerline{$^{21}$Institute for Theoretical and Experimental Physics,        
                   Moscow, Russia}                                            
\centerline{$^{22}$Moscow State University, Moscow, Russia}                   
\centerline{$^{23}$Institute for High Energy Physics, Protvino, Russia}       
\centerline{$^{24}$Lancaster University, Lancaster, United Kingdom}           
\centerline{$^{25}$University of Arizona, Tucson, Arizona 85721}              
\centerline{$^{26}$Lawrence Berkeley National Laboratory and University of    
                  California, Berkeley, California 94720}                     
\centerline{$^{27}$University of California, Davis, California 95616}         
\centerline{$^{28}$California State University, Fresno, California 93740}     
\centerline{$^{29}$University of California, Irvine, California 92697}        
\centerline{$^{30}$University of California, Riverside, California 92521}     
\centerline{$^{31}$Florida State University, Tallahassee, Florida 32306}      
\centerline{$^{32}$University of Hawaii, Honolulu, Hawaii 96822}              
\centerline{$^{33}$Fermi National Accelerator Laboratory, Batavia,            
                   Illinois 60510}                                            
\centerline{$^{34}$University of Illinois at Chicago, Chicago,                
                   Illinois 60607}                                            
\centerline{$^{35}$Northern Illinois University, DeKalb, Illinois 60115}      
\centerline{$^{36}$Northwestern University, Evanston, Illinois 60208}         
\centerline{$^{37}$Indiana University, Bloomington, Indiana 47405}            
\centerline{$^{38}$University of Notre Dame, Notre Dame, Indiana 46556}       
\centerline{$^{39}$Iowa State University, Ames, Iowa 50011}                   
\centerline{$^{40}$University of Kansas, Lawrence, Kansas 66045}              
\centerline{$^{41}$Kansas State University, Manhattan, Kansas 66506}          
\centerline{$^{42}$Louisiana Tech University, Ruston, Louisiana 71272}        
\centerline{$^{43}$University of Maryland, College Park, Maryland 20742}      
\centerline{$^{44}$Boston University, Boston, Massachusetts 02215}            
\centerline{$^{45}$Northeastern University, Boston, Massachusetts 02115}      
\centerline{$^{46}$University of Michigan, Ann Arbor, Michigan 48109}         
\centerline{$^{47}$Michigan State University, East Lansing, Michigan 48824}   
\centerline{$^{48}$University of Nebraska, Lincoln, Nebraska 68588}           
\centerline{$^{49}$Columbia University, New York, New York 10027}             
\centerline{$^{50}$New York University, New York, New York 10003}             
\centerline{$^{51}$University of Rochester, Rochester, New York 14627}        
\centerline{$^{52}$State University of New York, Stony Brook,                 
                   New York 11794}                                            
\centerline{$^{53}$Brookhaven National Laboratory, Upton, New York 11973}     
\centerline{$^{54}$Langston University, Langston, Oklahoma 73050}             
\centerline{$^{55}$University of Oklahoma, Norman, Oklahoma 73019}            
\centerline{$^{56}$Brown University, Providence, Rhode Island 02912}          
\centerline{$^{57}$University of Texas, Arlington, Texas 76019}               
\centerline{$^{58}$Texas A\&M University, College Station, Texas 77843}       
\centerline{$^{59}$Rice University, Houston, Texas 77005}                     
\centerline{$^{60}$University of Washington, Seattle, Washington 98195}       
}                                                                             
%end                                                                          

\date{\today}
\maketitle
%------------------------------------------------------------------------
\begin{abstract}
We report on a study of the ratio
	of inclusive three-jet 
	to inclusive two-jet production cross sections
	as a function of total transverse energy
	in $p\bar{p}$ collisions
	at a center-of-mass energy $\sqrt{s} = 1.8$ TeV,
	using data
	collected with the D\O\ detector during the 1992--1993 run of
    	the Fermilab Tevatron Collider.
The measurements 
	are used to 
	deduce preferred renormalization scales
	in perturbative ${\cal O}(\alpha_s^3)$
	QCD calculations
	in modeling soft-jet emission.
\end{abstract}
%------------------------------------------------------------------------
\twocolumn	%% take out for style committee
%------------------------------------------------------------------------
%   \section{Introduction}

A primary manifestation of Quantum Chromodynamics (QCD) 
	in $p\bar{p}$ collisions
	at a high center-of-mass energy $(\sqrt{s} = 1.8$ TeV)
	is the production of jets
	with large transverse momenta.
Typically, the hard interaction of parton constituents
    of a proton and an antiproton produce two hard back-to-back jets.
However, a fraction of the time,
	additional jets are also produced.
In the absence of an all-orders QCD calculation,
	jet production rates as a function of jet energy
	are predicted by fixed-order calculations in perturbative QCD (pQCD).
%--These calculations 
%--	depend on the choice of renormalization and factorization scales.
In this paper,
	we investigate the dependence of these calculations on 
	the choice of parton distribution functions (pdf)
	and particularly renormalization and factorization scales 

We examine the ratio of inclusive three-jet production 
	to inclusive two-jet production,
	which reflects the rate of 
	gluon emission
	in QCD jet production processes.
%--This ratio minimizes the sensitivity to 
%--	systematic uncertainties in the absolute cross section
%--	and can be compared 
%--	to pQCD predictions
%--	to probe the accuracy of the calculation
%--	specifically with regard to the choice of renormalization scale.
A three jet cross section explicitly offers the opportunity 
	to investigate a scale difference at a secondary vertex.
%%--This is the lowest order at which the possibility of 
%%--	such scale dependence is guaranteed.
Taking the ratio reduces systematic uncertainties.

Although this issue has inherent theoretical interest,
	it is also important because
	QCD multijet production is frequently
	a background to rare processes:
phenomenologically confirmed prescriptions
	for renormalization scales
	are essential for predicting background rates 
	and for
    	designing efficient triggering schemes for rare processes
	at future colliders~\cite{summers}.
Lastly, when higher order QCD calculations 
	become available,
	this ratio 
	may be useful for 
	providing another accurate measure of 
	the strong coupling constant $\alpha_s.$

%   \section{The Experiment}

The data used in this analysis, 
	corresponding to an integrated luminosity 
	of $\approx 10~{\text{pb}}^{-1},$
	were recorded during the 1992--1993 Tevatron collider run.
The D\O\ detector is described in detail elsewhere~\cite{D0-DETECTOR}.
Jet detection primarily utilizes 
	the uranium-liquid argon calorimeters,
	which have full coverage for pseudorapidity $|\eta| \le 4$
	where $\eta = -{\ln}[{\tan}(\theta/2)]$
	and $\theta$ is the polar angle relative to the 
	direction of the proton beam.
Initial event selection occurred in two hardware trigger
	stages and a software stage.
The first hardware trigger selected an inelastic $p\bar{p}$ collision 
	as indicated by signals from trigger hodoscopes 
	located near the beams 
	on either side of the interaction region.
The next stage required transverse energy above a preset threshold
	in calorimeter towers of 
  	$0.2 \times 0.2$ in $\Delta\eta\times\Delta\phi,$
	where $\phi$ is the azimuthal angle.
Selected events were digitized and sent to an array of processors.
Jet candidates were then reconstructed with a cone algorithm
	and the event recorded if any 
	jet transverse energy $(E_T)$
	exceeded a specified threshold.
Five such inclusive triggers
    had thresholds  of $20, 30, 50, 85,$ and $115~{\text{GeV}}.$

Jets were reconstructed offline 
    using an iterative fixed-cone algorithm 
    with a cone radius ${\cal R}=0.7$ in $\eta-\phi$ space.
The $E_T$ of each jet was corrected for 
	effects due to
	the underlying event,
	additional interactions,
	noise from uranium decay,
	the fraction of particle energy 
	deposited outside of the reconstruction cone,
	detector uniformity, and 
	detector hadronic response.
A discussion of the jet algorithm, 
	energy scale calibration and resolution
	can be found in Refs. \cite{levan,jetprd,jetnim}.

We measure the ratio of the inclusive three-jet 
	to the inclusive two-jet cross section
$$
R_{32} = \frac{\sigma_3}{\sigma_2} =
\frac{\sigma(p\overline{p} \rightarrow \it{n}~{\text{jets}} + X; \it{n} \geq 3)}
     {\sigma(p\overline{p} \rightarrow \it{m}~{\text{jets}} + X; \it{m} \geq 2)}
$$
as a  function of
    the scalar sum of jet transverse energies
    $(H_T = \sum E_T^{\text{jet}}).$
The measurement is performed for four distinct sets of selection criteria
	for all jets in the event:
$E_T$ thresholds of $20,30,$ or $40$ GeV for $|\eta_{\text{jet}}|<3,$
and $E_T>20$ GeV for $|\eta_{\text{jet}}|< 2$.
%-- add this
Three thresholds were chosen to study threshold dependence,
	and the minimum threshold was chosen to maximize statistics
	for which jet reconstruction efficiency was nearly 100\%.
%%--
Both in the data analysis and in the QCD calculation,
	a jet contributes to $H_T$ and to the jet multiplicity if
	it passes all selection criteria and
	satisfies the $E_T$ and $\eta_{\text{jet}}$ 
	requirements.

\begin{figure}
\epsfxsize=3.375in
\epsffile{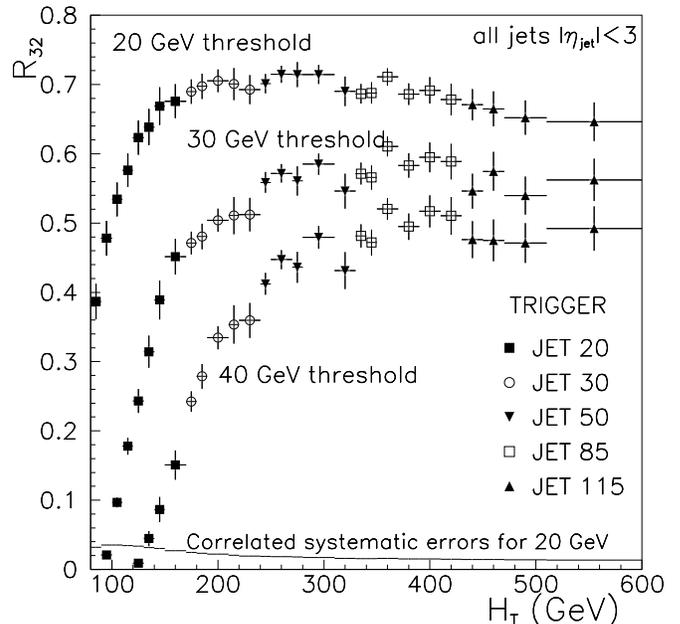}
\caption{The ratio $R_{32}$ as a function of $H_T$
	for $E_T$ thresholds of $20,$ $30,$ and $40$ GeV
	$(|\eta_{\text{jet}}|<3).$
    Error bars indicate statistical 
	and uncorrelated systematic uncertainties,
    while the distribution at the bottom
	shows the correlated systematic uncertainty
    for the $20$ GeV threshold.
\label{f:ratio}}
\end{figure}

Figure \ref{f:ratio}
    shows the ratio $R_{32}$ as a function of $H_T$
    for $E_T$ thresholds of $20,$ $30,$ and $40$ GeV
	for $|\eta_{\text{jet}}|<3.$
%----------------------------------------------------------------------
%%%%%%%%FIG1
%----------------------------------------------------------------------
The five trigger samples listed in the figure
	contribute in separate regions of $H_T,$ as indicated
	by the symbols.
The distribution at the bottom of the figure
	shows the correlated systematic uncertainties
	for the $20$ GeV threshold.
This uncertainty is
	the maximum offset in the ratio obtained by a 
	one standard deviation change
	in the correction to the jet energy scale.
Error bars indicate statistical 
%-- add this
	uncertainties (calculated using the appropriate 
	binomial prescription for a statistically correlated ratio)
%-- 
	as well as uncorrelated systematic uncertainties
	arising from all selection criteria.
Table~\ref{t:numbers}
	displays the measurements in bins of $H_T,$
	showing the 
	uncorrelated 
	and correlated uncertainties
    	for the four selection criteria.
%%%%%%%% TABLE numbers %%%%%%%%%%%%%%%%%%%%%%%%%%%%%%%%%%%%%%%%%%
\begin{table*}[hbt]
\begin{center}
\begin{tabular}{ccccc} 
\multicolumn{1}{c}  { $H_T$ Range   }   &
    \multicolumn{4}{c}  { $R_{32} \pm $ uncorrelated $\pm$ correlated uncertainty } \\ \cline{2-5}
\multicolumn{1}{c}  { (GeV)     }   &
    \multicolumn{1}{c}  { $E_T \ge 20$ GeV, $|\eta|<2$ } &
    \multicolumn{1}{c}  { $E_T \ge 20$ GeV, $|\eta|<3$ } &
    \multicolumn{1}{c}  { $E_T \ge 30$ GeV, $|\eta|<3$  } &
    \multicolumn{1}{c}  { $E_T \ge 40$ GeV, $|\eta|<3$  } \\ \hline
$ 80- 90$ &$0.315\pm.019\pm.029$ &$0.387\pm.025\pm.032$ &                     &                     \\
$ 90-100$ &$0.408\pm.018\pm.031$ &$0.478\pm.025\pm.035$ &$0.021\pm.003\pm.011$ &                     \\
$100-110$ &$0.444\pm.018\pm.029$ &$0.534\pm.024\pm.035$ &$0.097\pm.007\pm.016$ &                     \\
$110-120$ &$0.496\pm.019\pm.027$ &$0.576\pm.024\pm.034$ &$0.178\pm.012\pm.018$ &                     \\
$120-130$ &$0.537\pm.021\pm.025$ &$0.623\pm.025\pm.034$ &$0.243\pm.017\pm.019$ &$0.009\pm.004\pm.001$ \\
$130-140$ &$0.562\pm.025\pm.023$ &$0.639\pm.026\pm.031$ &$0.314\pm.023\pm.019$ &$0.045\pm.011\pm.004$ \\
$140-150$ &$0.579\pm.027\pm.021$ &$0.669\pm.027\pm.030$ &$0.389\pm.028\pm.019$ &$0.086\pm.018\pm.007$ \\
$150-170$ &$0.581\pm.025\pm.018$ &$0.676\pm.024\pm.027$ &$0.452\pm.026\pm.018$ &$0.151\pm.021\pm.010$ \\
$170-180$ &$0.616\pm.016\pm.017$ &$0.690\pm.018\pm.025$ &$0.471\pm.016\pm.017$ &$0.242\pm.015\pm.013$ \\
$180-190$ &$0.623\pm.017\pm.016$ &$0.698\pm.018\pm.023$ &$0.481\pm.018\pm.016$ &$0.279\pm.018\pm.013$ \\
$190-210$ &$0.612\pm.016\pm.014$ &$0.706\pm.016\pm.022$ &$0.504\pm.017\pm.016$ &$0.334\pm.016\pm.014$ \\
$210-220$ &$0.631\pm.025\pm.014$ &$0.701\pm.023\pm.021$ &$0.511\pm.027\pm.016$ &$0.354\pm.028\pm.013$ \\
$220-240$ &$0.615\pm.023\pm.013$ &$0.693\pm.021\pm.019$ &$0.512\pm.024\pm.015$ &$0.359\pm.025\pm.011$ \\
$240-250$ &$0.638\pm.014\pm.013$ &$0.701\pm.014\pm.019$ &$0.559\pm.015\pm.017$ &$0.412\pm.016\pm.012$ \\
$250-270$ &$0.656\pm.012\pm.012$ &$0.715\pm.012\pm.018$ &$0.572\pm.013\pm.017$ &$0.447\pm.014\pm.012$ \\
$270-280$ &$0.651\pm.020\pm.012$ &$0.714\pm.018\pm.018$ &$0.561\pm.021\pm.017$ &$0.436\pm.022\pm.011$ \\
$280-310$ &$0.661\pm.015\pm.012$ &$0.715\pm.014\pm.017$ &$0.585\pm.015\pm.017$ &$0.479\pm.016\pm.011$ \\
$310-330$ &$0.635\pm.023\pm.011$ &$0.690\pm.021\pm.016$ &$0.546\pm.025\pm.016$ &$0.431\pm.027\pm.009$ \\
$330-340$ &$0.653\pm.015\pm.011$ &$0.687\pm.014\pm.016$ &$0.571\pm.016\pm.017$ &$0.481\pm.017\pm.010$ \\
$340-350$ &$0.650\pm.016\pm.011$ &$0.688\pm.015\pm.016$ &$0.566\pm.017\pm.017$ &$0.472\pm.019\pm.010$ \\
$350-370$ &$0.669\pm.014\pm.011$ &$0.711\pm.013\pm.016$ &$0.611\pm.014\pm.018$ &$0.521\pm.016\pm.011$ \\
$370-390$ &$0.653\pm.017\pm.011$ &$0.686\pm.016\pm.015$ &$0.583\pm.018\pm.017$ &$0.495\pm.019\pm.010$ \\
$390-410$ &$0.653\pm.020\pm.011$ &$0.692\pm.019\pm.015$ &$0.595\pm.022\pm.018$ &$0.517\pm.023\pm.011$ \\
$410-430$ &$0.652\pm.024\pm.011$ &$0.678\pm.023\pm.015$ &$0.589\pm.026\pm.017$ &$0.510\pm.027\pm.010$ \\
$430-450$ &$0.643\pm.023\pm.011$ &$0.671\pm.023\pm.014$ &$0.546\pm.025\pm.016$ &$0.476\pm.027\pm.010$ \\
$450-470$ &$0.640\pm.027\pm.011$ &$0.665\pm.025\pm.014$ &$0.575\pm.028\pm.017$ &$0.475\pm.030\pm.010$ \\
$470-510$ &$0.634\pm.026\pm.011$ &$0.652\pm.025\pm.014$ &$0.540\pm.027\pm.016$ &$0.471\pm.029\pm.010$ \\
$510-600$ &$0.624\pm.029\pm.011$ &$0.646\pm.028\pm.014$ &$0.562\pm.031\pm.017$ &$0.492\pm.032\pm.010$
\end{tabular}
\caption{Values of $R_{32}$ with their uncorrelated and correlated
uncertainties for the indicated jet $E_T$ threshold and $\eta_{\text{jet}}$ criteria.
Uncorrelated uncertainties include statistical 
	and uncorrelated systematic uncertainties added in 
	quadrature.
\label{t:numbers}}
\end{center}
\end{table*}
%%%%%%%% TABLE numbers %%%%%%%%%%%%%%%%%%%%%%%%%%%%%%%%%%%%%%%%%%
%----------------------------------------------------------------------
%----------------------------------------------------------------------
%   \section{Theory Comparison }

{\sc jetrad}~\cite{jetrad}
    is a next-to-leading-order Monte Carlo generator
    for describing inclusive multijet production.
The generated 2-jet and 3-jet events are inclusive,
	and therefore the ratio of these cross sections 
	should be equivalent
    to the measured $R_{32}.$
CTEQ4M~\cite{cteq4m} pdf
	are used in the {\sc jetrad} simulations.
The jet finding algorithm in {\sc jetrad} approximates the algorithm
    used in D\O\ data reconstruction.
Jets generated by {\sc jetrad} are 
	individually 
	smeared according to
    known detector resolutions.
Two partons are combined if they are
    within ${\cal R}_{\text{sep}}=1.3{\cal R}$, 
	as \mbox{motivated} by the
    separation of jets in the data~\cite{rsep}
	and, just as in the data,
a jet is included
	if its $E_T$
	and $\eta_{\text{jet}}$ meet the chosen selection criteria.

In pQCD,
    the renormalization procedure introduces a mass scale $\mu_R$
	to control
	ultraviolet divergences
    	in the calculations.
A factorization scale $\mu_F$,
	introduced to handle infrared divergences,
	is assumed to be equal to $\mu_R$  
	in all predictions described in this paper.
QCD provides the 
	evolution
	of $\alpha_s$ with $\mu_R$,
	but not its absolute scale.
Unless otherwise indicated,
	the renormalization scale $\mu_R = \lambda H_T$
	will be used for the production 
	of the two leading jets,
where the constant $\lambda$, the coefficient of the hard scale,
    	will have a nominal value of $0.3,$
	but will be allowed to vary as described below.
To study the possibility of having a 
	different scale for 
	the production of additional jets,
    the renormalization scale of the third jet
	is varied from
    $\mu_R^{(3)} = \lambda H_T$ (same as for the leading jets)
    to a scale proportional to the $E_T$
    of the third jet $\mu_R^{(3)} \propto E_T^{(3)}.$
Also,
	a scale proportional to the maximum jet transverse energy 
	$(E_T^{\text{max}})$ is studied, 
	as this is a standard form used
	for comparisons of {\sc jetrad} 
	to measured jet cross sections.

\begin{figure}
\epsfxsize=3.375in
\epsffile{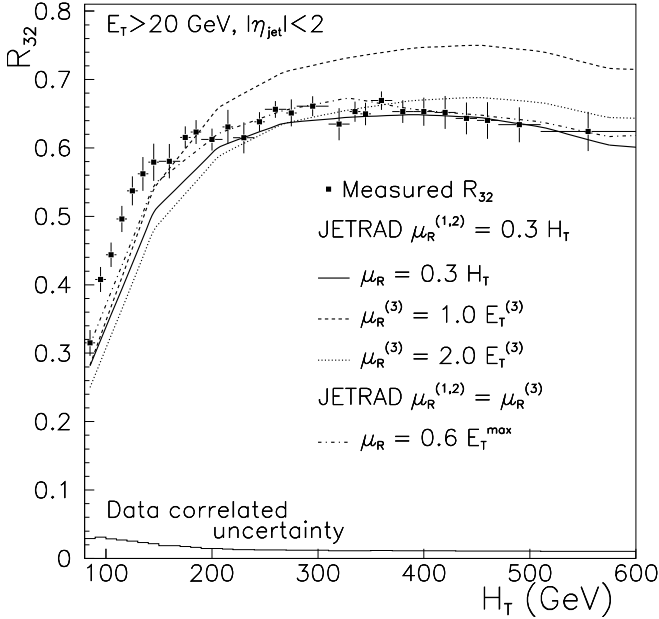}
\caption{ The ratio $R_{32}$ as a function of $H_T,$
    	requiring jet $E_T>20$ GeV and $|\eta_{\text{jet}}|<2.$
    Error bars indicate statistical and 
	uncorrelated systematic uncertainties,
	while the histogram at the bottom
	shows the correlated systematic uncertainty.
    The four smoothed distributions show the {\sc jetrad} prediction
	for the renormalization scales indicated in the legend.
\label{f:r32_all}}
\end{figure}

%--Figure \ref{f:r32_all}
%--	shows the measured $R_{32}$ as a function of
%--	$H_T$ for jet $E_T>20$ GeV and $|\eta_{\text{jet}}|<2,$
%--	the data selection criteria
%--	offering reduced uncertainties in the measurement
%--	while preserving maximum sensitivity 
%--	to scale in the {\sc jetrad} prediction.
Figure \ref{f:r32_all} shows the measured $R_{32}$ as a function of
	$H_T$ for jet $E_T>20$ GeV and $|\eta_{\text{jet}}|<2.$
The 20 GeV threshold has good sensitivity 
	to scale in the {\sc jetrad} prediction
	and has reduced statistical uncertainty.
The central rapidity region has the best understood
	jet energy uncertainties and correlations.
%----------------------------------------------------------------------
%%%%%%%%FIG2
%----------------------------------------------------------------------
The plot contains four smoothed distributions corresponding to
	{\sc jetrad} predictions for  
	the following renormalization prescriptions
	(shown for $\lambda=0.3$):
\begin{itemize}
\item $\mu_R= \lambda H_T$ for the two leading jets,
  \begin{itemize}
  \item $\mu_R= \lambda H_T$ also for the third jet (solid)
  \item $\mu_R^{(3)} = E_T^{(3)}$ for the third jet (dashed)
  \item $\mu_R^{(3)} = 2 E_T^{(3)}$ for the third jet (dotted)
  \end{itemize}
\item $\mu_R = 0.6 E_T^{\text{max}}$ for all jets (dash-dot).
\end{itemize}
All predictions demonstrate
    the same qualitative behavior as the $R_{32}$ measurement, that is,
    a rapid rise below $H_T=200$ GeV 
	(associated with the kinematic threshold), 
	a leveling off, then
  	a slight drop at highest $H_T$
	(associated with the reduced phase space
	for additional radiation for high $E_T$ jets).
Although {\sc jetrad} predictions for the ratio 
	are found to be insensitive to
	the choice of pdf,
	they do depend on the choice of ${\cal R}_{\text{sep}}.$
Allowing ${\cal R}_{\text{sep}}$ 
	to vary such that
  	neighboring 
  	jets are all merged or all split
  	causes a $3\%$ decrease 
  	or increase in the ratio,
  	respectively, 
  	with only a slight effect on the shape 
  	of the distribution in $H_T$.

\begin{figure}
\epsfxsize=3.375in
\epsffile{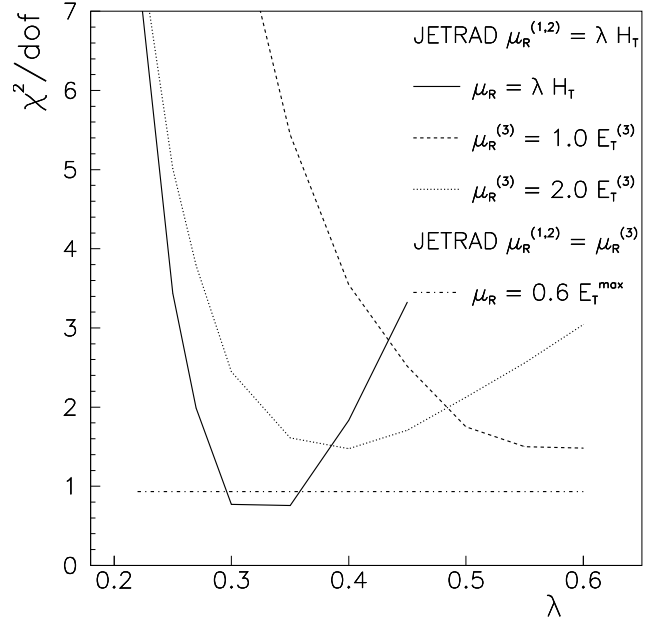}
\caption{$\chi^2/$dof as a function of $\lambda,$ comparing
    data to {\sc jetrad} predictions for several 
    renormalization prescriptions
    for the $E_T>20$ GeV, $|\eta_{\text{jet}}|<2$ selection criteria.
\label{f:chi2}}
\end{figure}

For a quantitative comparison,
	we use a 
	$\chi^2$ covariance technique, defining
$$
    \chi^2 = \sum_{j=1}^{n}      \sum_{i=1}^{n}
    (D_i - T_i)
	C_{ij}^{-1}
	(D_j - T_j)
$$
where $D_i$ and $T_i$ represent the ${i^{\text{th}}}$ data and theory
    element, respectively, and
    ${\bf C^{-1}}$ is the inverse of the covariance matrix.
This matrix incorporates 
    uncorrelated uncertainties in the measurement and
    statistical uncertainties in the simulation,
    with correlated uncertainties 
	included for the absolute jet energy in the data
	and for the uncertainty from resolution smearing 
	in {\sc jetrad} (not shown explicitly in Fig. \ref{f:r32_all}).
Although some of the predictions
	do not visually overlap with the data,
	acceptable agreement is found for some scales
	because of the strong point-to-point correlations 
	of the data uncertainties  
	which are taken into account in the $\chi^2.$
Figure \ref{f:chi2} 
    shows the $\chi^2$ per degree of freedom $(\chi^2/$dof)
    as a function of the parameter $\lambda,$
    for the $E_T>20$ GeV, $|\eta_{\text{jet}}|<2$ selection criteria.
%----------------------------------------------------------------------
%%%%%%%%FIG3
%----------------------------------------------------------------------
The degrees-of-freedom equal the number of data points (28).
The horizontal line indicates the $\chi^2/$dof obtained 
	using the $\lambda$ independent scale $\mu_R = 0.6 E_T^{\text{max}}$
	for all jets.
This scale yields good agreement with measurement 
	(probability $p>57\%)$ 
	for the $E_T>20$ GeV criteria,
	but the $\chi^2$ rises (and the corresponding probabilities decrease)
	for the higher $E_T$ thresholds (not shown).

For $\lambda$-dependent scales, 
	the best fit is specified by the $\lambda$ 
	that minimizes the $\chi^2.$
The scales proportional to $E_T^{(3)}$ for the third jet
	do not provide a good fit 
	$(p<5\%)$ 
	for any $\lambda,$
	as seen in Fig. \ref{f:chi2}.
While there is fair agreement
	in the wider region of pseudorapidity $|\eta_{\text{jet}}|<3$
	for certain regions of $\lambda$ (not shown), 
	these do not correspond to the same values
	for different $E_T$ thresholds,
	making the applicability of this scale prescription
	unsuitable for predicting 
	production rates for additional jets.

The {\sc jetrad} prediction assuming a scale 
	$\mu_R = \lambda H_T$ 
	provides the best description of the data
	for $\lambda$ between $0.30$ and $0.35~(p>80\%).$
Moreover, the $\chi^2$ is also minimized 
	in the $\lambda\approx0.30$ region
	for the other selection criteria (not shown)
	making this scale choice the most robust 
	of all the $\mu_R$ scales studied.

%----------------------------------------------------------------------
%   \section{Conclusions}

In conclusion,
    we have measured the ratio of the 
	inclusive three-jet  
	to  the inclusive two-jet cross  section
    as a  function of total scalar transverse energy $H_T$
	and compared the results to {\sc jetrad} predictions.
The greatest sensitivity to
	the choice of renormalization scale is for 
	the lowest $E_T$ threshold of $20$ GeV.
Although no prediction accurately describes
	the ratio through the kinematic threshold region,
a single $\mu_R$ scale assumption in the calculation
	for all jets 
	is found to adequately describe 
	the rate of additional jet emission
	when correlated uncertainties are accounted for
	in a $\chi^2$ comparison.
Specifically, a scale of $\mu_R= \lambda H_T$ for all jets,
	where $\lambda= 0.3,$
	yields a prediction consistent with the measurement
	for all jet-selection criteria examined.
A scale of  $\mu_R= 0.6 E_T^{\text{max}}$ for all jets
	also provides a sufficient description 
	at the lowest jet $E_T$ threshold.
The introduction of additional scales 
	does not significantly improve agreement with the data.
	
%----------------------------------------------------------------------
%   \section*{Acknowledgments}

We thank David Summers, Dieter Zeppenfeld, and Walter Giele for
stimulating and helpful discussions.
% Acknowledgement_paragraph.tex
%
We thank the staffs at Fermilab and at collaborating institutions 
for contributions to this work, and acknowledge support from the 
Department of Energy and National Science Foundation (USA),  
Commissariat  \` a L'Energie Atomique and
CNRS/Institut National de Physique Nucl\'eaire et 
de Physique des Particules (France), 
Ministry for Science and Technology and Ministry for Atomic 
   Energy (Russia),
CAPES and CNPq (Brazil),
Departments of Atomic Energy and Science and Education (India),
Colciencias (Colombia),
CONACyT (Mexico),
Ministry of Education and KOSEF (Korea),
CONICET and UBACyT (Argentina),
A.P. Sloan Foundation,
and the A. von Humboldt Foundation.
%

%------------------------------------------------------------------------
%------------------------------------------------------------------------

%------------------------------------------------------------------------
\end{document}